\documentclass[10pt,letterpaper]{article}
\usepackage{opex3}
\usepackage{cite}
\usepackage{graphicx}
\usepackage{epstopdf}

\usepackage{ae} 

\begin{document}

\title{Mode instability thresholds for Tm-doped fiber amplifiers pumped at 790~nm}

\author{Arlee V. Smith$^*$ and Jesse J. Smith}

\address{AS-Photonics, LLC, 6916 Montgomery Blvd. NE, Suite B8, Albuquerque, NM 87109 USA}

\email{$^*$arlee.smith@as-photonics.com}

\begin{abstract}
We use a detailed numerical model of stimulated thermal Rayleigh scattering to compute mode instability thresholds in Tm$^{3+}$-doped fiber amplifiers. The fiber amplifies 2040~nm light using a 790~nm pump. The cross-relaxation process is strong, permitting power efficiencies of 60\%.   The predicted instability thresholds are compared with those in similar Yb$^{3+}$-doped fiber amplifiers with 976~nm pump and 1060~nm signal, and are found to be higher, even though the heat load is much higher in Tm-doped amplifiers. The higher threshold in the Tm-doped fiber is attributed to its longer signal wavelength, and to stronger gain saturation, due in part to cross-relaxation heating.
\end{abstract}


\section{Introduction}

Kilowatt signal power levels have been achieved in Yb$^{3+}$-doped fiber amplifiers operating near 1.0~$\mu$m and in Tm$^{3+}$-doped fiber amplifiers operating near 2.0~$\mu$m. One obstacle encountered in attempting to scale single-transverse-mode outputs from Yb-doped fibers to still higher powers is a thermally induced mode instability\cite{Eidam:2011}. Such mode instability has not yet been reported for Tm-doped fiber. However, considering its thermal origin and the fact that the heat deposited in Tm-doped fiber is more than three times higher than in Yb-doped fiber, an important question is whether such instability is expected for 790~nm pumped Tm-doped fiber amplifiers, and if so, at what threshold power. In earlier reports we developed a model of stimulated thermal Rayleigh scattering (STRS) to explain  mode instability in Yb-doped fiber amplifiers\cite{Smith:2011,Smith:2013a,Smith:2014a}. Here we extend our STRS model to Tm-doped fiber to confirm the likelihood of mode instability and to predict threshold signal powers.

Thulium-doped amplifiers have several attractions. Like Yb-doped fibers, they are broadly tunable so they support ultra short pulse amplification. Atmospheric transmission is good near 2040~nm where these fibers operate well at high power. Their relatively long signal wavelength makes it easier to design fibers with large mode area, leading to better suppression of stimulated Brillouin scatter (SBS). The SBS linewidth is reduced by approximately four relative to a Yb-doped fiber, so suppression by temperature tuning the SBS Stokes shift or broadening the signal linewidth are also more effective than in Yb-doped fibers\cite{Goodno:2011}. The longer signal wavelength also means the self focusing power limit, which is proportional to ($\lambda^2/nn_2$), is approximately four times higher than in Yb-doped fiber. Additionally, the pump absorption band at 790~nm enables pumping by high power, high efficiency AlGaAs pump lasers.

High power Tm amplifiers pumped at 790~nm have been demonstrated by Ehrenreich {\it et al.}\cite{Ehrenreich:2010} who achieved 1 kW at 2045~nm in a 12 m long 20:400 (core:pump cladding diameters) Tm-doped silicate fiber with numerical aperture (NA) of 0.08 (V=3.1). Their slope efficiency was 53\%. Goodno {\it et al.}\cite{Goodno:2009} demonstrated 608~W at 2040~nm, pumping with 790~nm light, to achieve 54\% power efficiency in 3.1~m of 25:400 fiber. High beam quality was demonstrated by their $M^2$ value of 1.05. Further, the phase stability of high power Tm-doped fibers was shown to be good enough for efficient coherent beam combining of multiple fiber outputs\cite{Gaida:2015,Goodno:2011}. However, the best power efficiency of Tm-doped amplifiers is only about 60\%, in contrast to 90\% for Yb-doped fiber pumped at 976~nm. %A further drawback is the doubled focal spot diameter on a target for 2~$\mu$m light relative to 1~$\mu$m light.

\section{Stimulated thermal Rayleigh scattering (STRS)}

First, we offer a brief review of the physics included in our model of STRS in fiber amplifiers. Mode instability is the degradation of output beam quality above a sharp power threshold\cite{Eidam:2011}. Assuming most of the signal seed light is injected into the fundamental transverse mode LP$_{01}$, below threshold the output remains mostly in LP$_{01}$. Above threshold a substantial fraction occupies higher order modes, especially LP$_{11}$. Because different transverse modes have different propagation constants, these two modes interfere to create a signal irradiance pattern that oscillates side-to-side along the length of the fiber. Pump light is preferentially absorbed in regions of higher signal irradiance where the population inversion is lower, and because a certain fraction of the absorbed pump is converted to heat, this creates a heating pattern that resembles the irradiance pattern. The heat pattern is converted to a similar temperature pattern, and the temperature pattern creates a refractive index pattern via the thermo-optic effect. If the interference pattern is stationary, there is little or no phase shift between the thermally-induced index pattern and the irradiance pattern, resulting in nearly zero net transfer of power between modes. However, if the light in the higher order mode is slightly detuned in frequency from LP$_{01}$, the irradiance pattern moves along the fiber - downstream for a red detuning and upstream for a blue detuning. The temperature pattern also moves but lags the irradiance pattern due to thermal diffusion. This lag produces the phase shift necessary to induce power transfer between modes. Red detuning of LP$_{11}$ leads to power transfer from LP$_{01}$ to LP$_{11}$. 

When light in LP$_{01}$ is transferred to LP$_{11}$ by the moving grating it experiences a frequency shift equal to the modal frequency offset due to a Doppler effect\cite{Smith:2011}, so the transferred light adds coherently to the light already in LP$_{11}$. The resulting exponential gain process can be categorized as near-forward stimulated thermal Rayleigh scattering. The frequency offset that maximizes the mode coupling is approximately the inverse of the thermal diffusion time across the fiber core. Diffusion times are approximately 0.1-1 ms, implying frequency detunings of approximately 1-10 kHz.

Our approach to modeling STRS is to develop the most general numerical model feasible. It includes all the physical effects just described\cite{Smith:2013a}. We use diffractive beam propagation of a time-periodic signal field. This field incorporates all transverse fiber modes (including lossy modes) and can include all offset frequencies that are harmonics of a selected principal offset frequency. All modes are assumed to have the same polarization. The time-dependent temperature profile is computed using a steady-periodic Green's function method\cite{Cole:2006}. Mode coupling occurs through inclusion of the thermally-induced change in the transverse refractive index profile used to compute the beam propagation. No analytic or semi-analytic expressions of mode coupling are needed, and coupling between all modes is included. We use this approach because it permits the most accurate modeling, and because it makes inclusion of various additional physical effects relatively straightforward. The cost of such a general numerical model is long run-times, but using the methods described in Ref.~\cite{Smith:2013a} we can model a few meters of fiber per hour on a consumer grade desktop computer. In the model results presented here we ignore pump and signal losses due to scattering or photodarkening, temperature dependence of the Tm ion absorption and emission cross-sections and the ion decay rates, and practical issues such as melting the fiber outer cladding. 

Alternative STRS models based on mode coupling theory have been published by other authors\cite{Hansen:2012,Ward:2012,Dong:2013,Tao:2015} but we will not review them here. An alternative beam propagation model that includes the same physics as ours but without the assumption of periodic behavior has also been developed\cite{Ward:2013}.

In comparing STRS gain in Yb- and Tm-doped fibers several factors must be considered. The first is the influence of the nearly doubled signal wavelengths in Tm-doped fiber. Another is the more complex process of heat generation in Tm-doped fiber. In Yb-doped fiber approximately 10\% of the absorbed pump light is converted to heat. If each pump photon creates one signal photon then the heat is the difference in photon energies of ($h [\nu_p - \nu_s]$) which is about 10\% of $h \nu_p$. In Tm-doped fiber if one pump photon created one signal photon then the heat would be approximately 60\% of $h \nu_p$ assuming 790~nm pump and 2040~nm signal, limiting the power efficiency to around 40\%. Measured power efficiencies are greater than 60\%, however, and this is attributed to a cross-relaxation process. Clearly this process must be included in an accurate Tm-doped fiber amplifier STRS model. As is true for Yb-doped fiber, population saturation has a strong impact on STRS thresholds in Tm-doped fiber. Population saturation leads to transverse heat profiles which more closely resemble a tophat than the fundamental mode signal irradiance which resembles a Gaussian. This reduces the antisymmetric part of the heat and pushes it toward the outer edge of the doped region. The result is reduced STRS coupling strength\cite{Smith:2013b,Hansen:2014,Tao:2015,Ward:2015}.

\section{Tm$^{3+}$ spectral properties}

\begin{figure}
\centering
\includegraphics[width=0.75\textwidth]{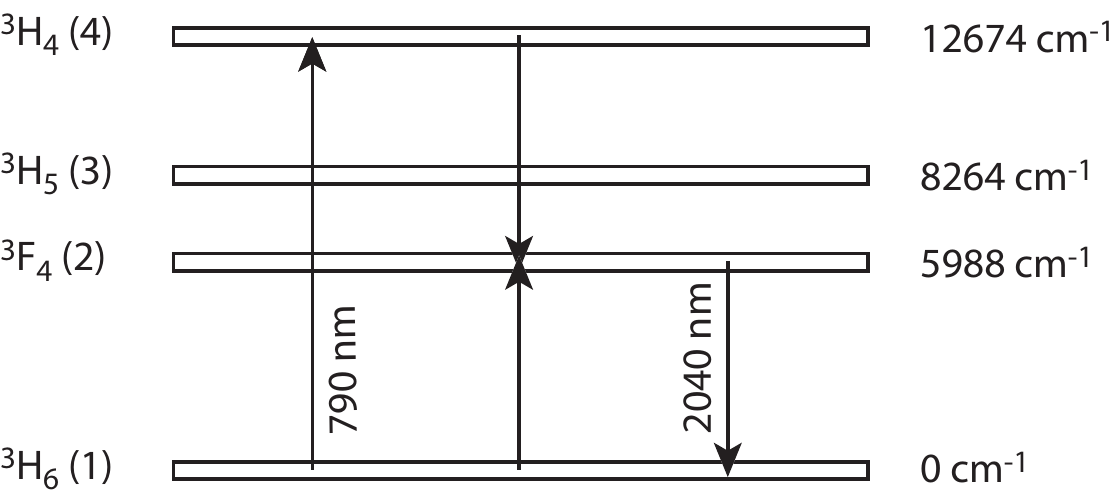}
\caption{Energy levels for Tm$^{3+}$ in silicate glass. The pump wavelength is 790~nm and the signal wavelength is 2040~nm. See Ref.~\cite{Peterka:2011}. The double arrows indicate the 2-for-1 cross-relaxation process.}
\label{fig.tm_levels}
\end{figure}

The energy level diagram for Tm$^{3+}$ in silicate glass is shown in Fig.~\ref{fig.tm_levels}. We will refer to the levels by number: $(1)=\,^3H_6$; $(2)=\,^3F_4$; $(3)=\,^3H_5$; $(4)=\,^3H_4$. We consider only 790~nm pumping and 2040~nm lasing. Level 4 is directly populated via pump absorption by level-1 ions, and gain occurs via stimulated emission from level 2 to level 1. One of the attractions of 790~nm pumping is a cross-relaxation process that combines a level-4 ion with a level-1 ion to produce two level-2 ions. This two-for-one process produces two ions in the upper laser level from the absorption of a single pump photon, and this allows the quantum efficiency to exceed unity. Measured power efficiencies exceeding 60\% have been demonstrated even though unit quantum efficiency corresponds to roughly 40\% power efficiency. For the cross-relaxation process to compete with nonradiative decay of the level-4 ions, the doping concentration of Tm ions should exceed the 2\% doping limit of silica. Therefore, Tm-doped fibers are usually based on silicate glass with several percent Al doping. The concentration of Tm can then be as high as 7\%, although 4-6\% seems to be the most common level\cite{Lee:2013,Lee:2015,Jackson:2003}.

\subsection{Population rate equations}

Assuming the population of level 3 is negligible because it decays quickly into level 2, and ignoring any excited state absorption, we write the population equations in the form
\begin{eqnarray}\label{eq.tm_pop1}
\frac{dn_1}{dt}&=&A_{21} n_2+A_{41}n_4+(\sigma_p^en_4-\sigma_p^an_1)I_p/h\nu_p\\\nonumber
&&-k_{41}^{22}N_{\circ}n_4n_1+k_{22}^{41}N_{\circ}n_2^2+(\sigma_s^en_2-\sigma_s^an_1)I_s/h\nu_s,\\\label{eq.tm_pop2}
\frac{dn_2}{dt}&=& A_{42} n_4-A_{21}n_2\\\nonumber
&&+2k_{41}^{22}N_{\circ}n_4n_1-2k_{22}^{41}N_{\circ}n_2^2-(\sigma_s^en_2-\sigma_s^an_1)I_s/h\nu_s,\\\label{eq.tm_pop4}
\frac{dn_4}{dt}&=&-A_{41}n_4-A_{42}n_4+(\sigma_p^an_1-\sigma_p^en_4)I_p/h\nu_p\\\nonumber
&&-k_{41}^{22}N_{\circ}n_4n_1+k_{22}^{41}N_{\circ}n_2^2,\\
\label{eq.tm_pop_total}1&=&n_1+n_2+n_4.
\end{eqnarray}
Here, $N_\circ$ is the Tm$^{3+}$ doping density, $A_{ij}$ is the rate of decay of level $i$ into level $j$, $\sigma_p^a$, $\sigma_p^e$, $\sigma_s^a$ and $\sigma_s^e$ are absorption and emission cross sections for the pump and signal, $I_p$ and $I_s$ are the irradiances of pump and signal, $k_{ij}^{mn}$ is the cross-relaxation coefficient for ions initially in levels $i$ and $j$ to transfer into levels $m$ and $n$, $h$ is Planck's constant, and $\nu$ is optical frequency of the pump or signal. In our model, we retain the $(x,y,t)$ variations of $I_i$, $n_i$, and the $(x,y)$ variations of $N_{\circ}$.

\subsection{Laser gain equations}

The laser gain equations take the form
\begin{equation}\label{eq.dis_dz}
\frac{dI_s}{dz}=N_{\circ}(\sigma_s^e n_2 - \sigma_s^a n_1)I_s,
\end{equation}
\begin{equation}\label{eq.dip_dz}
\frac{dI_p^\pm}{dz}=\pm N_{\circ}(\sigma_p^e n_4 - \sigma_p^a n_1)I_p^\pm,
\end{equation}
where $I_p^+$ is the  co-propagating pump, and $I_p^-$ is the counter-propagating pump.

\subsection{Steady state populations}

If we assume the signal and pump irradiances vary slowly in time, we can set the time derivatives equal to zero in Eqs.~(\ref{eq.tm_pop1})-(\ref{eq.tm_pop4}) and solve for the steady state populations in terms of the irradiances, cross sections, decay and cross-relaxation rates. The steady state population equations are 
\begin{eqnarray}\label{eq.tm_pop1ss}
0&=&A_{21} n_2+A_{41}n_4+(\sigma_p^en_4-\sigma_p^an_1)I_p/h\nu_p\\\nonumber
&&-k_{41}^{22}N_{\circ}n_4n_1+k_{22}^{41}N_{\circ}n_2^2+(\sigma_s^en_2-\sigma_s^an_1)I_s/h\nu_s,\\\label{eq.tm_pop2ss}
0&=& A_{42} n_4-A_{21}n_2+2k_{41}^{22}N_{\circ}n_4n_1-2k_{22}^{41}N_{\circ}n_2^2\\\nonumber
&&-(\sigma_s^en_2-\sigma_s^an_1)I_s/h\nu_s,\\\label{eq.tm_pop4ss}
0&=&-A_{41}n_4-A_{42}n_4+(\sigma_p^an_1-\sigma_p^en_4)I_p/h\nu_p\\\nonumber
&&-k_{41}^{22}N_{\circ}n_4n_1+k_{22}^{41}N_{\circ}n_2^2,\\\label{eq.tm_pop1234ss}
1&=&n_1+n_2+n_4.
\end{eqnarray}
We solve these equations for the population fractions $n_1$, $n_2$, and $n_4$ by substitution and use of the quadratic formula. We verified our solutions by integrating the population rate equations Eqs.~(\ref{eq.tm_pop1})-(\ref{eq.tm_pop4}) in time until steady state is obtained.

\subsection{Heat deposition}

The density of power left behind at each point in the core due to the difference in absorbed pump light and stimulated signal emission can be written
\begin{equation}\label{eq.power_lost}
Q=-\frac{dI_p^+}{dz}+\frac{dI_p^-}{dz}-\frac{dI_s}{dz}.
\end{equation}
Expressions for the derivatives of pump and signal irradiance on the right side of this equation are given by Eqs.~(\ref{eq.dis_dz}) and (\ref{eq.dip_dz}) in terms of absorption and emission cross sections. In the steady state, those expressions can be evaluated with the help of Eqs.~(\ref{eq.tm_pop2ss}) and (\ref{eq.tm_pop4ss}), and substituted into Eq. (\ref{eq.power_lost}) to arrive at
\begin{eqnarray}\label{eq.heat2}
Q &=&N_{\circ}n_4(A_{42}[h\nu_p-h\nu_s]+A_{41}h\nu_p)+N_{\circ}n_2A_{21}h\nu_s\\\nonumber
&&+N_{\circ}^2(k_{41}^{22}n_1n_4-k_{22}^{41}n_2^2)(h\nu_p-2h\nu_s).
\end{eqnarray}
Not all of that left-behind power actually turns into heat since some of it is radiated away. We account for the portion that does become heat by using $A_{ij}^{\prime}$ in place of $A_{ij}$ where the primed value is somewhat smaller than the unprimed one, to find the equation for heat deposition
\begin{eqnarray}\label{eq.heat3}
{Q^\prime}&=&N_{\circ}n_4(A_{42}^{\prime}[h\nu_p-h\nu_s]+A_{41}^{\prime}h\nu_p)+N_{\circ}n_2A_{21}^{\prime}h\nu_s\\\nonumber
&&+N_{\circ}^2(k_{41}^{22}n_1n_4-k_{22}^{41}n_2^2)(h\nu_p-2h\nu_s).
\end{eqnarray}
Heating due to additional absorption processes such as signal absorption due to photodarkening, or signal absorption by core impurities can be added if desired.

\section{Choices of doping density, decay rates, and cross sections}

\subsection{Tm$^{3+}$ doping levels}

Accurate modeling using the population, gain, and heat equations requires good knowledge of the decay rates, cross-relaxation rates, and doping levels. These are each discussed below, and the values used in our model are listed in Table~\ref{tab.comparison_yb_tm}. As mentioned earlier, high power Tm-doped fibers usually use aluminum silicate glass, and many of the reports of high power Tm amplifiers cite a doping density of approximately 4\%. The doping level is critical because the cross-relaxation rate is linear in $N_{\circ}$. We use $N_{\circ}=5\times 10^{26}$/m$^3$.

\subsection{Signal emission cross section, $\sigma_s^e$}

\begin{figure}[htb]
\centering
\includegraphics[width=0.85\textwidth]{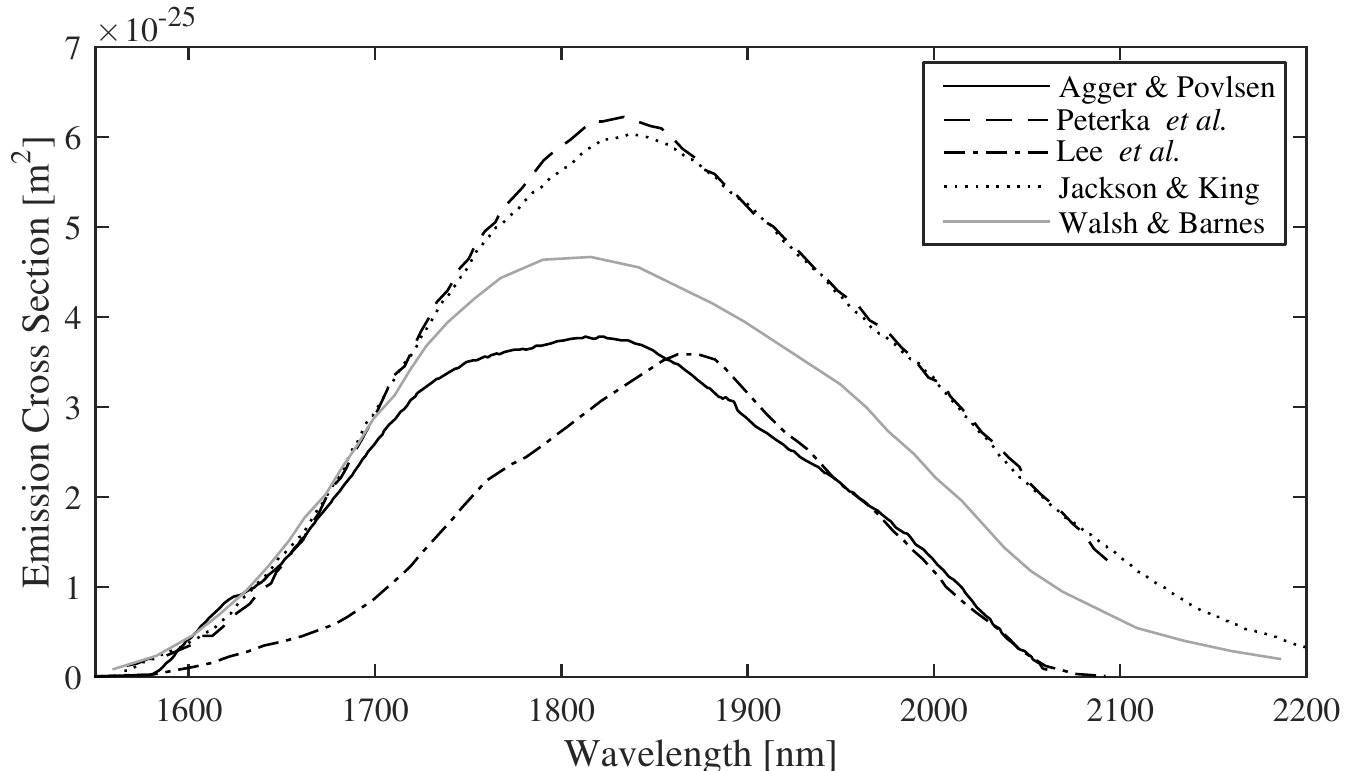}
\caption{\label{fig.sigma_s_e}Signal emission cross sections for the $^3H_6-^3F_4$ (1$\leftrightarrow$2) transition in thulium-doped aluminum silicate fiber, from Refs.~\cite{Agger:2006, Peterka:2011, Lee:2015, Jackson:1999, Walsh:2004}.}
\end{figure}

In Fig.~\ref{fig.sigma_s_e} we show several reported signal emission cross sections for thulium-doped silicate fiber. One is from Agger and Povlsen\cite{Agger:2006}. Their method for calculating the cross sections from observed differences in input and output light, without knowledge of the doping density, is described in \cite{Agger:2006}. Their  doping density is then estimated as $8.4\times 10^{25}$m$^{-3}$. Their signal emission cross section is scaled using the radiative lifetime of level 2, which they take to be 6.0 ms. A second cross section is from Jackson and King\cite{Jackson:1999}. Their peak cross section is $6\times 10^{-25}$ m$^2$, 50\% larger than Agger and Povlsen's. Their curve also extends further to the red than Agger and Povlsen's. Lee {\it et al.}\cite{Lee:2015} show a third measured emission cross section, in this case in silicate glass  with $N_{\circ}=8.35\times 10^{26}$m$^{-3}$. Walsh and Barnes\cite{Walsh:2004} present a fourth cross section, for silica. Their peak value is $4.6\times 10^{-25}$ m$^2$. Turri {\it et al.}\cite{Turri:2008} also give $\sigma_s^e$ for Tm in silica (not shown). Their peak value at 300 K is $4\times 10^{-25}$ m$^2$. Peterka {\it et al.}\cite{Peterka:2011} report a sixth cross section for Tm in silicate glass that closely resembles that of Jackson and King.

\begin{figure}[htb]
\centering
\includegraphics[width=0.85\textwidth]{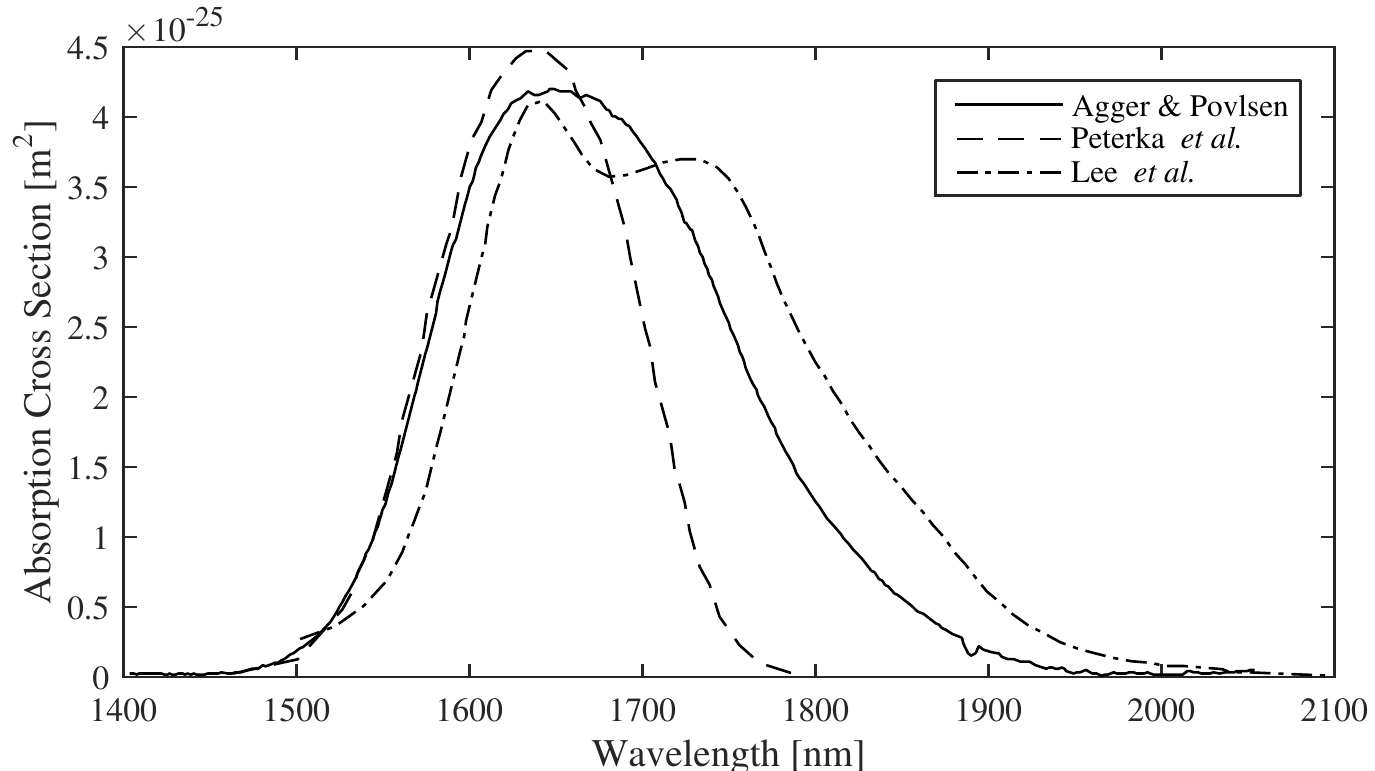}
\caption{\label{fig.sigma_s_a}Signal absorption cross sections for the $^3H_6-^3F_4$ (1$\leftrightarrow$2) transition in thulium-doped aluminum silicate fiber, from Refs.~\cite{Agger:2006, Peterka:2011, Lee:2015}.}
\end{figure}

Considering all these reported cross sections we use $\sigma_s^e=1.5\times 10^{-25}$ m$^2$ for our signal wavelength of 2040~nm. Note that increasing temperature enhances the blue side of the emission cross section curve at the expense of the rest of the curve, as illustrated by Turri {\it et al.}\cite{Turri:2008}. In our model we use a fixed cross section, independent of the core temperature. Since the laser wavelength is on the red edge of the emission cross section curve, the change with temperature is probably modest.

\subsection{Signal absorption cross section, $\sigma_s^a$}
 
In Fig.~\ref{fig.sigma_s_a} we display reported signal absorption curves. One is from Agger and Povlsen\cite{Agger:2006}. Another is from Jackson and King\cite{Jackson:1999}. Their curve is nearly identical to that of Peterka {\it et al.}\cite{Peterka:2011} and is not shown in Fig.~\ref{fig.sigma_s_a}. The Jackson and King and Peterka {\it et al.} curves do not extend as far to the red as Agger and Povlsen's, cutting off at 1800~nm. Their peak value is nearly the same as Agger and Povlsen's. Lee {\it et al.}\cite{Lee:2015} also report a measured absorption cross section for silicate glass  with $N_{\circ}=8.35\times 10^{26}$m$^{-3}$. Note that increasing the temperature should enhance the red side of the absorption cross section curve and this increases reabsorption of the signal. Based on these absorption curves we use $\sigma_s^a=0.1\times 10^{-25}$ m$^2$ at 2040~nm.

\subsection{Pump absorption cross section near 790~nm, $\sigma_p^a$}

\begin{figure}
\centering
\includegraphics[width=0.85\textwidth]{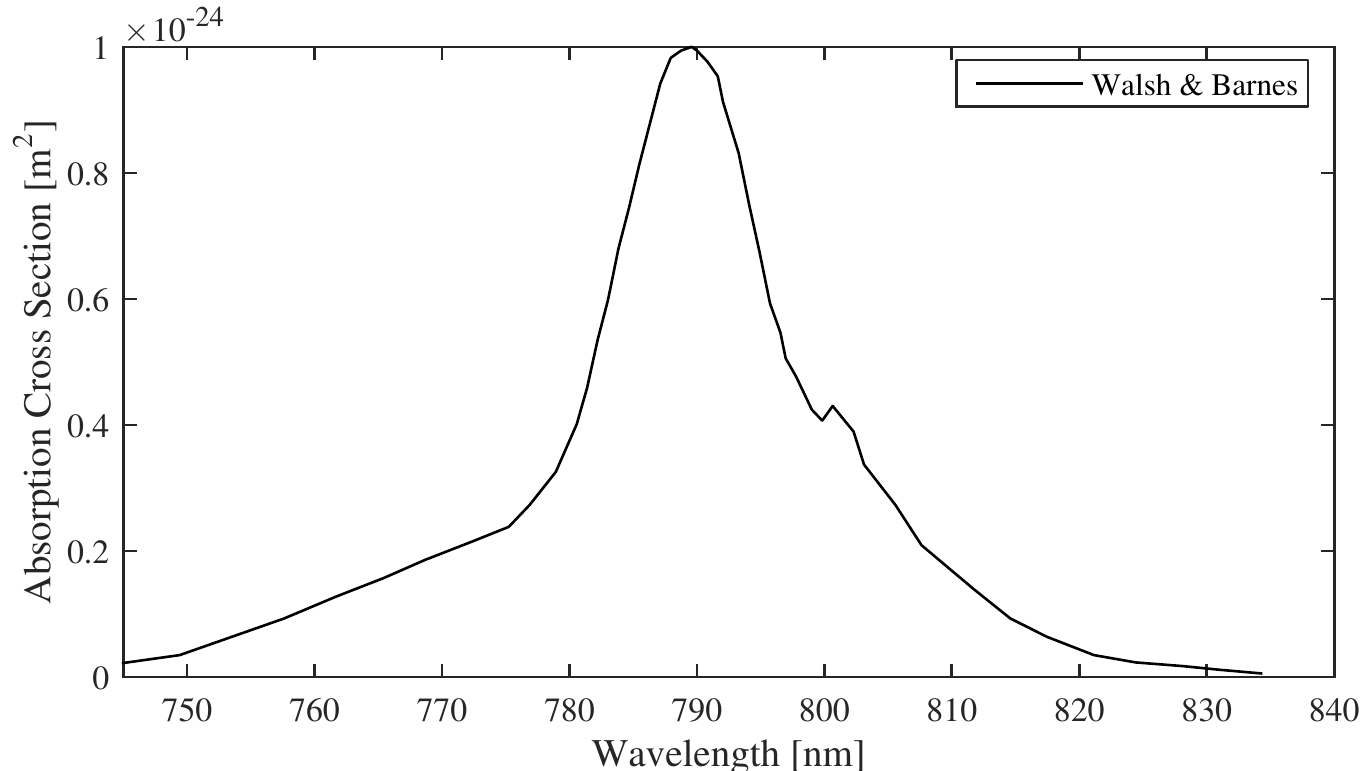}
\caption{Walsh and Barnes \cite{Walsh:2004} absorption cross section for Tm$^{3+}$ ions in silica fiber.}
\label{fig.pump_absn_cross_section}
\end{figure}

Jackson and King\cite{Jackson:1999} report a pump absorption curve (see Fig.~3 of Ref.~\cite{Jackson:1999}), with peak cross section of $8.5\times 10^{-25}$ m$^2$ at 800~nm. Walsh and Barnes\cite{Walsh:2004} show a plot of pump absorption cross section of Tm in silica which peaks near 790~nm at a value of $10\times 10^{-25}$ m$^2$ and has a width (FWHM) of approximately 15~nm. This curve is reproduced here as Fig.~\ref{fig.pump_absn_cross_section}. Peterka {\it et al.}\cite{Peterka:2011} also report a peak absorption of $8.5\times 10^{-25}$ m$^2$ at 789~nm. Turri {\it et al.}\cite{Turri:2008} show a maximum cross section of only $4.5\times 10^{-25}$ m$^2$ for Tm:silica.

We take the peak cross section to be $10 \times 10^{-25}$ m$^2$, and use as an effective cross section $\sigma_p^a=6\times 10^{-25}$ m$^2$ to account for a likely pump linewidth of several~nm.

\subsection{Pump emission cross section near 800~nm, $\sigma_p^e$}

Peterka {\it et al.}\cite{Peterka:2011} present uncalibrated curves of absorption and emission cross sections for the level 1 to level 4 transition near 800~nm. The two curves have similar linewidths, with the emission curve shifted about 20~nm to the red of the absorption curve. The emission cross section near 790~nm is lower than the absorption cross section by a factor of 10-15. We use $\sigma_p^e = 0.5 \times 10^{-25}$ m$^2$.

\subsection{Cross-relaxation coefficient $k_{41}^{22}$}

This coefficient sets the rate for populating level 2 with two ions at the expense of one ion from level 4 and one ion from level 1. Jackson and King\cite{Jackson:1999} list the value as $18\times 10^{-23}$ m$^3$~s$^{-1}$. Moulton {\it et al.}\cite{Moulton:2009} do not produce a number for this coefficient, but they find when $N_{\circ}=3\times 10^{26}$m$^{-3}$ about 0.75 of the decay from level 4 is due to cross relaxation. The remaining 0.25 is due to decay with a 45 $\mu$s lifetime measured at low doping levels where cross relaxation can be ignored. This implies $k_{41}^{22}=22\times 10^{-23}$ m$^3$~s$^{-1}$, in fair agreement with Jackson and King. We use $k_{41}^{22}=20\times 10^{-23}$ m$^3$~s$^{-1}$ in our model.

\subsection{Cross relaxation coefficient $k_{22}^{41}$}

In thermodynamic equilibrium this rate is related to $k_{41}^{22}$ by detailed balance which can be expressed
\begin{equation}
g_1g_4\exp[-E_4/kT]k_{41}^{22}=g_2g_2\exp[-2E_2/kT]k_{22}^{41},
\end{equation}
where $E_i$ is the energy of level $i$, and $g_i$ its degeneracy. Solving for $k_{22}^{41}$ gives
\begin{equation}
k_{22}^{41}=k_{41}^{22}\frac{g_1g_4}{g_2^2}\exp[(2E_2-E_4)/kT].
\end{equation}
Using $E_2=6000$ cm$^{-1}$ ($1.19\times 10^{-19}$ J), $E_4=12700$ cm$^{-1}$ ($2.52\times 10^{-19}$ J), and $T=300$ K ($kT=4.14\times 10^{-21}$ J) gives
\begin{equation}
k_{22}^{41}=0.05\;k_{41}^{22}.
\end{equation}

The ratio of cross relaxation rates is quite temperature sensitive, with the $k_{22}^{41}$ rate increasing relative to the $k_{41}^{22}$ rate as the temperature rises. Both rates probably increase with temperature as well. We use a ratio of 10\% instead of 5\% because the core temperature of a high power fiber amplifier is quite high. We use $k_{22}^{41}=2.0\times 10^{-23}$ m$^3$~s$^{-1}$ in our model.

\subsection{Rates $A_{21}$, $A_{21}^{\prime}$}

Agger and Povlsen\cite{Agger:2006} state the radiative life time of level 2 is 6.0 ms. They measured a lifetime of only 650 $\mu$s, so the decay is mostly nonradiative, with $A_{21}=1.58\times 10^3$~s$^{-1}$ and $A_{21}^{\prime}=0.9\;A_{21}=1.42\times 10^3$~s$^{-1}$. Jackson and King\cite{Jackson:1999} find the lifetime of level 2 is 335 $\mu$s, corresponding to a rate of $A_{21}=3.0\times 10^3$~s$^{-1}$. Assuming a 6.0 ms radiative lifetime, $A_{21}^{\prime}=0.94\;A_{21}=2.8\times 10^3$~s$^{-1}$. Lee {\it et al.}\cite{Lee:2013} report the lifetime of level 2 in heavily doped glass is 635 $\mu$s, making $A_{21}=1.57\times 10^3$~s$^{-1}$, in close agreement with Agger and Povlsen. Subtracting the assumed radiative rate of 167s$^{-1}$ gives $A_{21}^{\prime}=1.4\times 10^3$~s$^{-1}$.

Moulton {\it et al.}\cite{Moulton:2009} report a double exponential decay from level 2, with decay rates of about 4000~s$^{-1}$ and 1550~s$^{-1}$. Most ions decayed at the slower rate; about 1/4 decayed at the fast rate. A value of 1850~s$^{-1}$ is a reasonable effective decay rate, making $A_{21}=1.85\times 10^3$~s$^{-1}$. Subtracting the assumed radiative rate of 167~s$^{-1}$ gives $A_{21}^{\prime}=1.7\times 10^3$~s$^{-1}$. Walsh and Barnes\cite{Walsh:2004} claim a radiative lifetime of 4.56 ms, computed using the Judd-Ofelt method. Peterka {\it et al.}\cite{Peterka:2011} state a lifetime in lightly doped fiber of 430 $\mu$s, or $A_{21}=2.3\times 10^3$~s$^{-1}$. 

The list of reported decay rates is thus ($A_{21}=1580, 3000, 1570, 1850, 2300$~s$^{-1}$), and we use $A_{21} = 2000$~s$^{-1}$. From this we subtract 167~s$^{-1}$ corresponding to a radiative lifetime of 6 ms to approximate $A_{21}^{\prime}$ as $A_{21}^{\prime}=1.8\times 10^3$~s$^{-1}$.

\subsection{Rates $A_{41}$, $A_{42}$, and $A_{42}^{\prime}$}

The lifetime of level 3 is short (7 ns according to Jackson and King\cite{Jackson:1999}) due to nonradiative decay to level 2, so decay from level 4 into level 3 can be considered equivalent to decay directly into level 2, allowing us to use an effective $A_{42}$ in place of $A_{42}$ and $A_{43}$.

Walsh and Barnes\cite{Walsh:2004} give a radiative lifetime for level 4 of 670 $\mu$s for a rate of 1490~s$^{-1}$. If we assign this decay entirely to level 1, we have $A_{41}=1490$~s$^{-1}$ and $A_{41}^{\prime}=0$. The measured lifetime is much shorter\cite{Peterka:2011,Moulton:2009}, only 45-58 $\mu$s, and it is probably mostly due to nonradiative decay to level 2 which would make $A_{42}=2\times 10^4$~s$^{-1}$ and $A_{42}^{\prime}=2\times 10^4$~s$^{-1}$. 

Peterka {\it et al.}\cite{Peterka:2011} report a 58 $\mu$s effective lifetime, but it is not single exponential. Based on their report, we estimate $(A_{41}+A_{42}=1.7\times 10^4$~s$^{-1}$$)$. Using $A_{41}=1490$~s$^{-1}$ gives $A_{42}=1.55\times 10^4$~s$^{-1}$.

For modeling we chose the values $A_{41}=1.5\times 10^3$~s$^{-1}$; $A_{41}^{\prime}=0$; $A_{42}=1.6\times 10^4$~s$^{-1}$; $A_{42}^{\prime}=1.6\times 10^4$~s$^{-1}$.

\begin{table}
\begin{center}
\begin{tabular}{|c|c|c|}
\hline
Parameter & Yb-doped &  Tm-doped\\
\hline
$\lambda_{p}$                &           976~nm        &      790~nm           \\
$\lambda_{s}$                &          1060~nm        &      2040~nm          \\
$L$                          &            4~m          &       4~m             \\
$P_{p}$                      &         500 \& 500 W    &      500 \& 500 W     \\
$d_{core}$                   &         25~$\mu$m       &       25~$\mu$m       \\
$d_{clad}$                   &         400~$\mu$m      &      400~$\mu$m       \\
$n_{core}$                   &         1.451           &       1.4537          \\ 
$n_{clad}$                   &         1.45            &       1.45            \\ 
$V$                          &         4.0             &      4.0              \\
$N_{dope}$               	 & $1.2\times 10^{26}$ m$^{-3}$ & $5.0\times 10^{26}$ m$^{-3}$\\
$d_{dope}$ 				 	 &          25~$\mu$m      &       25~$\mu$m       \\
$\sigma_p^a$             &$2.47\times 10^{-24}$ m$^2$ 	&  $6.0\times 10^{-25}$ m$^2$ \\
$\sigma_p^e$             &$2.44\times 10^{-24}$ m$^2$ 	&  $0.5\times 10^{-25}$ m$^2$ \\
$\sigma_s^a$             &$6.0\times 10^{-27}$ m$^2$ 	&  $0.1\times 10^{-25}$ m$^2$ \\
$\sigma_s^e$             &$3.58\times 10^{-25}$ m$^2$ 	&  $1.5\times 10^{-25}$ m$^2$ \\
LP$_{01}$ Power 		 &  6 W                         & 6 W                   \\
$A_{21}(A_{21}^{\prime})$   & $1.1\times 10^3(1.1\times 10^3)$~s$^{-1}$ & $2.0\times 10^3(1.8\times 10^3)$~s$^{-1}$ \\
$A_{41}(A_{41}^{\prime})$            &        &  $1.5\times 10^3(0)$~s$^{-1}$ \\
$A_{42}(A_{42}^{\prime})$            &   & $1.6\times 10^4(1.6\times 10^4)$~s$^{-1}$\\
$k_{41}^{22}$                &                         & $20\times 10^{-23}$ m$^3$~s$^{-1}$ \\
$k_{22}^{41}$                &                         & $2.0\times 10^{-23}$ m$^3$~s$^{-1}$ \\
$\rho$ 	&	2201 kg/m$^3$ & 2201 kg/m$^3$ \\
$\kappa$ & 1.38 W/m$\cdot$K & 1.38 W/m$\cdot$K \\
${\rm d} n/{\rm d}T$ & $1.2 \times 10^{-5}$ K$^{-1}$ & $1.2 \times 10^{-5}$ K$^{-1}$ \\
$C$ & 703 J/kg$\cdot$K & 703 J/kg$\cdot$K \\
\hline
\end{tabular}
\caption{Parameters used in comparing similar Yb- and Tm-doped fiber amplifiers.}
\label{tab.comparison_yb_tm}
\end{center}
\end{table}

\subsection{Steady state approximation}

Our model runs fastest when we can use the steady state values of the populations. That means the populations must converge to their steady state values in less time than the inverse of the frequency offset associated with highest STRS gain. This time is approximately equal to the thermal diffusion time across the core\cite{Smith:2011}. For the 25~$\mu$m diameter core that we model below the thermal diffusion time is about 175 $\mu$s corresponding to a frequency of approximately 6 kHz. Therefore, if the settling time for the populations is less than 60 $\mu$s for all combinations of signal and pump power, the steady state approximation should be safe.

To check whether this condition is met we integrate the population equations \mbox{Eqs.~(\ref{eq.tm_pop1})-(\ref{eq.tm_pop_total})} for different combinations of signal (790~nm) and pump (2040~nm) power to find the slowest convergence time likely to be encountered in a high power amplifier. Convergence always occurs in less than 54~$\mu$s, justifying the steady state approximation.

\section{Comparing Yb and Tm modeled mode instability thresholds}

\begin{table}[htb]
\begin{center}
\begin{tabular}{|c|c|c|c|c|c|c|}
\hline
 & \multicolumn{2}{|c|}{Co-pumped} & \multicolumn{2}{|c|}{Counter-pumped} & \multicolumn{2}{|c|}{Bidirectionally-pumped} \\
	       & Yb   & Tm   & Yb   & Tm   & Yb   & Tm  \\\hline
Signal out & 1451 & 2046 & 2233 & 3947 & 2235 & 3210 \\
Pump in    & 1600 & 3100 & 2410 & 6585 & 2503 & 4906 \\
Pump abs.  & 1570 & 3005 & 2362 & 6007 & 2456 & 4686 \\
Left-behind& 127  & 959  & 191  & 2066 & 198  & 1483 \\ 
\hline
\end{tabular}
\caption{Powers (in watts) at mode instability thresholds computed for similar Yb- and Tm-doped fiber amplifiers with the parameters given in Table~\ref{tab.comparison_yb_tm}. Input signal power is 6~W. Left-behind power is the difference between the absorbed pump power and the increase in signal power. Most of this left-behind power is heat.}
\label{tab.tm_yb_threshold_sig_out_pump_in}
\end{center}
\end{table}

\begin{figure}[htb]
\centering
\includegraphics[width=0.85\textwidth]{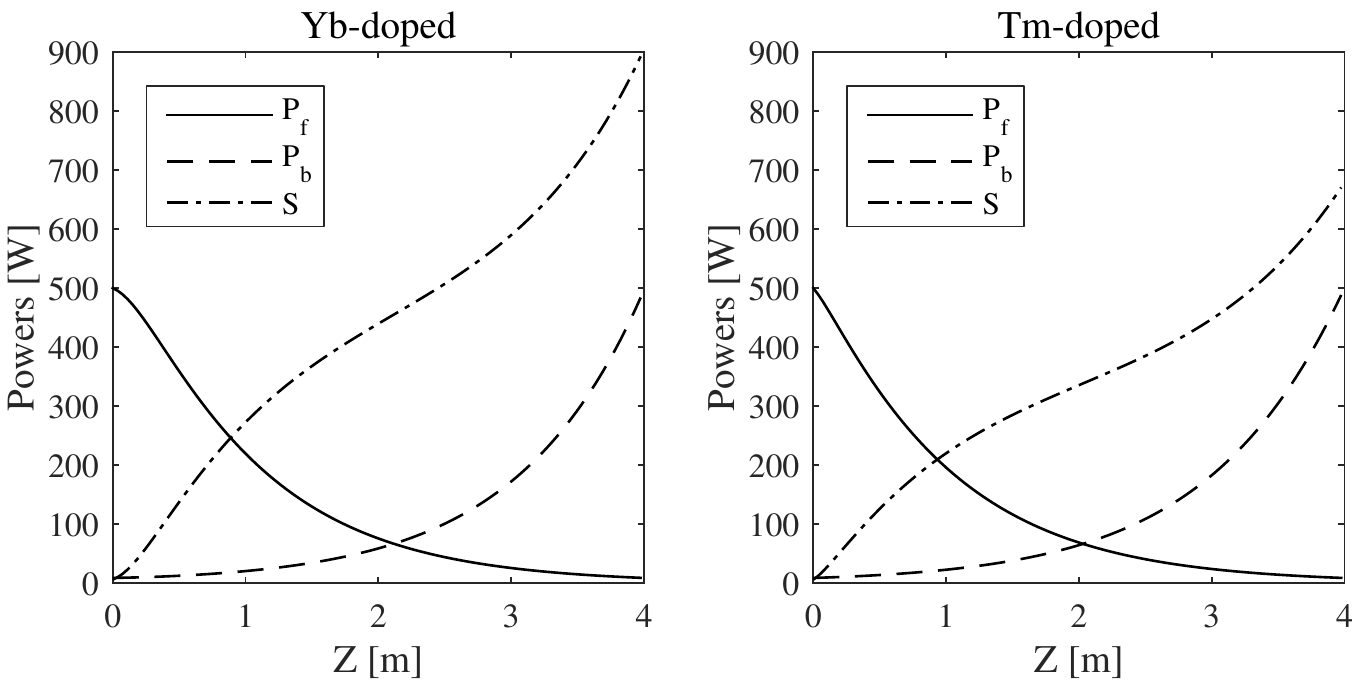}
\caption{Power curves generated by our simplified model for Yb- and Tm-doped fibers in the bidirectionally-pumped case with total pump power of 1000 W, and the parameters given in Table~\ref{tab.comparison_yb_tm}. Approximately 100~W of power is lost from the pump and signal in the Yb-doped fiber and 325~W in the Tm-doped fiber.}
\label{fig.simple_yb_tm_comparison_500W}
\end{figure}

It will be interesting to compare thresholds in Tm-doped fibers with those of Yb-doped fibers. To make this comparison more meaningful, we make the fibers as similar as possible by using the parameters listed in Table~\ref{tab.comparison_yb_tm}. Both fibers have the same core and cladding sizes, length, and signal seed powers. They also have the same V-number which means the mode profiles are the same despite the different signal wavelengths. The LP$_{01}$-LP$_{11}$ intermodal beat length in the Yb-doped fiber is 92\% greater than the Tm-doped fiber's, which closely matches the ratio of signal wavelengths. However, this difference in beat lengths has little effect on the STRS threshold. 

Before applying our full STRS model to compute thresholds, we use a simplified model to compare powers along the fibers using transversely-resolved populations to calculate signal and pump gains and losses\cite{Smith:2011b}. This simplified model does not include diffractive beam propagation, thermal lensing or STRS gain, and only LP$_{01}$ is populated. We apply this model to a bidirectional pumping configuration with 500~W of pump entering each end of the fiber. \mbox{Figure \ref{fig.simple_yb_tm_comparison_500W}} compares the evolution of pump and signal powers for the Yb- and Tm-doped fibers. An important point is that we have adjusted the doping densities so the pump absorption is nearly the same in the two cases. However, the signal output power is 893~W in the Yb-doped case, and 681~W in the Tm-doped case. This means the power lost to heat in Yb-doped fiber is approximately 10\% of absorbed pump power, while in Tm-doped fiber it is approximately 30\%. If total heat were the only important factor in mode instability, this would suggest Tm-doped fiber should have a threshold power one third that of Yb-doped fiber.

\begin{figure}[htb]
\centering
\includegraphics[width=0.85\textwidth]{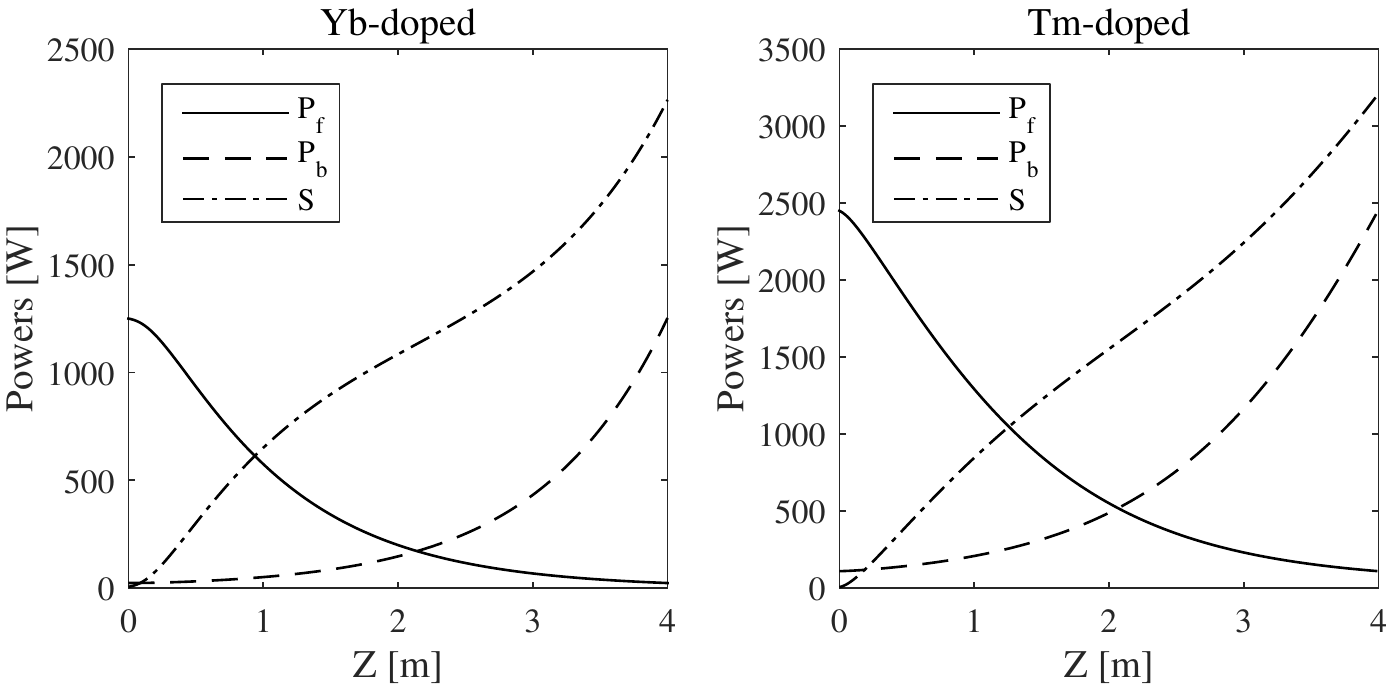}
\caption{Power curves generated by our full STRS model for operation at threshold of bidirectionally pumped Yb- and Tm-doped fiber amplifiers using the parameters listed in Table~\ref{tab.comparison_yb_tm}.}
\label{fig.yb_tm_bipump_threshold_power_curve}
\end{figure}

To model mode instability thresholds for these two cases, we use our full STRS model (detailed in Refs.~\cite{Smith:2013a,Smith:2014a}) which automatically includes thermal lensing. We use the thermal properties of silica even though the properties of the silicate host glass used for Tm-doped fibers probably differ somewhat. The fibers are unbent, step-index fibers with uniform doping across the full core. We define threshold as the signal output power at which LP$_{11}$ content reaches 1\%. 

We seed only the LP$_{01}$ and LP$_{11}$ modes because LP$_{11}$ has the highest STRS gain. We use a single frequency stand-in for seeding at the quantum noise level, which is a minimum seed level and yields the highest thresholds. More realistic seed power can be estimated for spontaneous thermal Rayleigh seeding\cite{Smith:2013c} or from pump or signal amplitude modulation\cite{Smith:2012a}, although thresholds are rather insensitive to this seed level. The signal is seeded with 6~W in LP$_{01}$ at zero frequency offset, and quantum noise level seeding of LP$_{11}$ is approximated by a seed of 10$^{-16}$~W red shifted by 4~kHz, where the model indicates the STRS gain is maximum. For bidirectionally-pumped cases, we apply half the total pump power to each end of the fiber. In this application of our model, we decompose the heat at each transverse point in the doped region into frequencies of 0 and 4~kHz which we use to compute the periodic temperature profiles using the Green's function \cite{Smith:2013a}. Previously, we showed this method provides thresholds that agree well with those from our model computed using higher spectral resolution\cite{Smith:2014a}.

\mbox{Figure \ref{fig.yb_tm_bipump_threshold_power_curve}} shows full STRS model results at threshold for bidirectionally-pumped Yb-doped and Tm-doped amplifiers. We have performed similar computations for copumped and counterpumped amplifiers as well. All the computed threshold powers are listed in Table~\ref{tab.tm_yb_threshold_sig_out_pump_in}. The Tm-doped fiber thresholds are universally substantially higher than Yb-doped fiber thresholds despite their much higher heat loads. The ratio of the left-behind power for the Tm-doped to Yb-doped fiber amplifiers at STRS threshold are in the range of 7.5-11, and almost all of that power is deposited in the fiber as heat.

\subsection{Thermal lensing}
\begin{figure}[htb]
\centering
\includegraphics[width=0.9\textwidth]{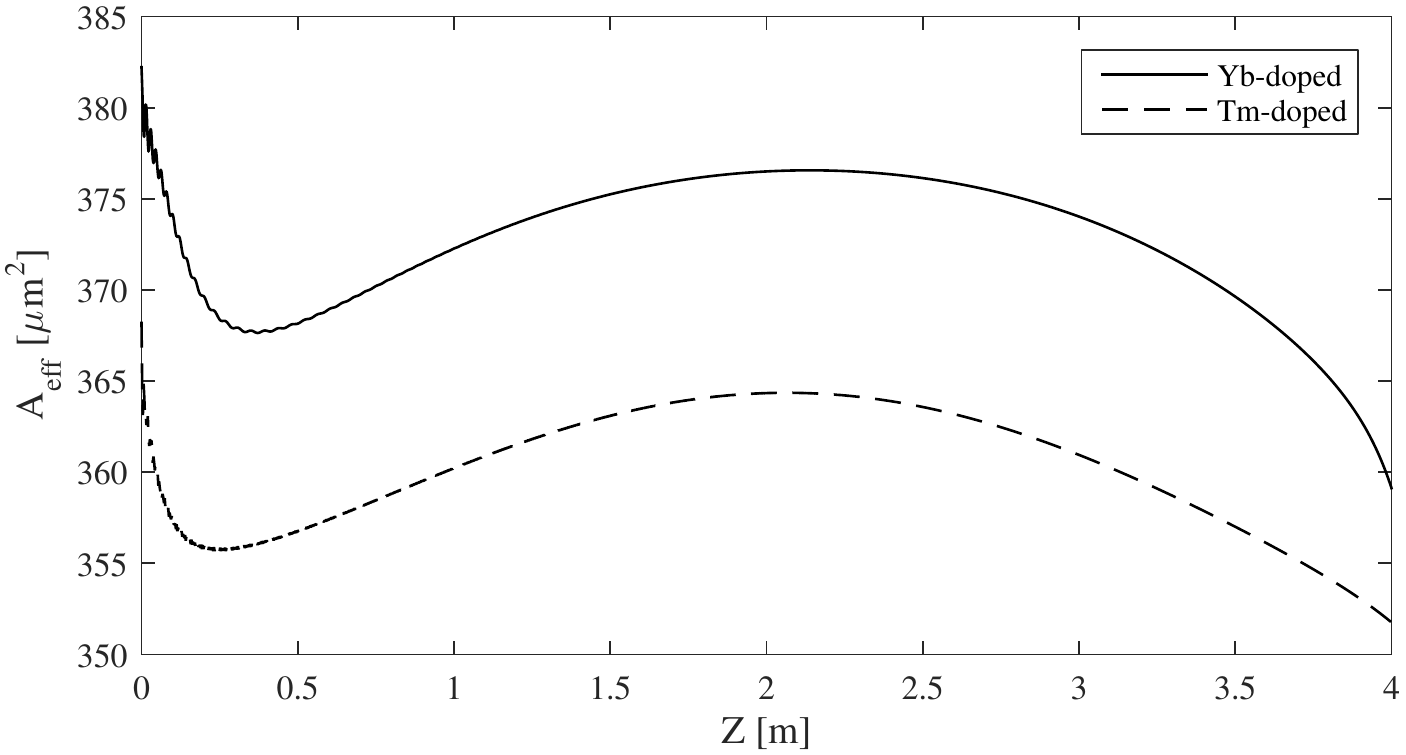}
\caption{Thermal lensing of fundamental mode for bidirectionally-pumped Yb-doped and Tm-doped fiber amplifiers near STRS threshold. These curves correspond to the powers plotted in Fig.~\ref{fig.yb_tm_bipump_threshold_power_curve}, and use parameters in Table~\ref{tab.comparison_yb_tm}. The thresholds are given in Table~\ref{tab.tm_yb_threshold_sig_out_pump_in}. The unlensed area is 380~$\mu$m$^2$.}
\label{fig.thermal_lensing}
\end{figure}

Given such high heat levels, one concern is the severity of thermal lensing. In Fig.~\ref{fig.thermal_lensing} we plot the effective area for bidirectionally pumped Yb-doped and Tm-doped fiber amplifiers near STRS threshold for comparison. We launch thermally-lensed modes to avoid strong oscillations in $A_{\rm eff}$. Even at the high powers involved, thermal lensing is modest. The longer signal wavelength of the Tm-doped amplifier helps to counteract thermal lensing, resulting in a decrease in effective area only about twice that of the Yb-doped fiber despite much higher heating.

\section{Discussion}
As Table~\ref{tab.tm_yb_threshold_sig_out_pump_in} shows, the ratio of heatloads at STRS threshold in Yb-doped and Tm-doped fibers is not a single fixed value, but varies from about 7.5 to 11. One contribution to the higher thresholds in Tm-doped fibers is the longer signal wavelength. The phase shift due to a fixed temperature rise over a given length of fiber is proportional to $k_\circ$ ($k_\circ = 2\pi/\lambda$) making the coupling between modes in Tm-doped fiber approximately half as large as in Yb-doped fiber. This $\lambda^{-1}$ scaling is evident in the coupling terms of coupled-mode theory for waveguides in general\cite{Marcuse:1974}, and in coupled-mode models of STRS in particular\cite{Hansen:2012,Dong:2013,Hansen:2014,Ward:2012,Tao:2015}. This factor accounts for approximately a factor of two in threshold heat loads, leaving a factor of three or so to be explained.

\begin{figure}[htb]
\centering
\includegraphics[width=0.9\textwidth]{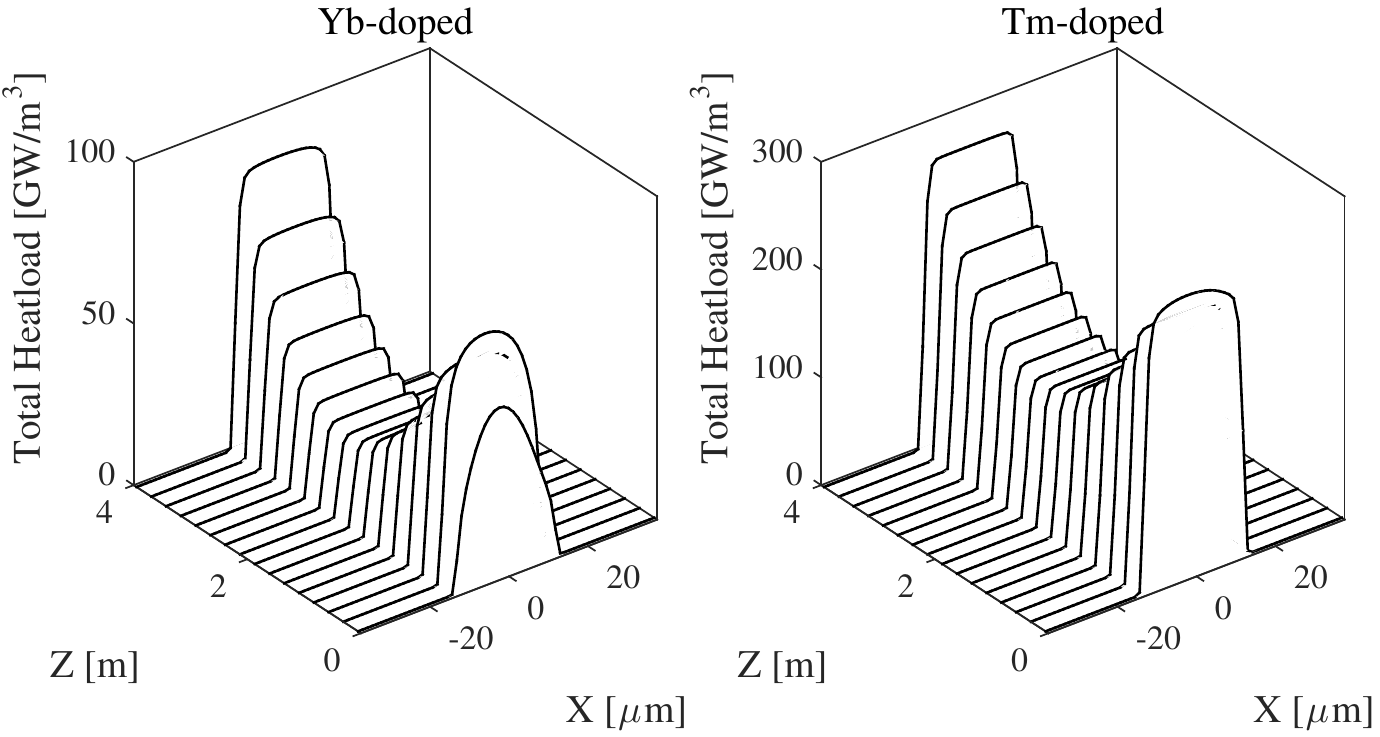}
\caption{Computed total heat profiles for cuts through the fiber center of the 25~$\mu$m diameter core Yb- and Tm-doped fibers of Table~\ref{tab.comparison_yb_tm} with each end of the fiber pumped with 500~pump. More than three times more heat is deposited in Tm-doped fiber than in Yb-doped fiber. Note the different vertical scales for the two plots.}
\label{fig.comparison_saturation_heat_yb_tm}
\end{figure}

To explore the causes of this remaining difference, we compare the heat distributions of Yb-doped and Tm-doped fibers in more detail. We use the same simplified model described above, and apply it to the same 1000 W bidirectional pumping case shown in Fig.~\ref{fig.simple_yb_tm_comparison_500W}. This pump power is well below threshold for both Yb- and Tm-doped fibers, but offers a direct comparison of heat profiles under similar conditions. We use the signal and pump powers shown in Fig.~\ref{fig.simple_yb_tm_comparison_500W} to calculate populations at several longitudinal ($z$) locations in the presence of additional LP$_{11}$ content, fixed at 1\% of signal power each $z$-location. The two modes are phased such that the modal interference effect is maximum. The modal interference produces a higher irradiance in the left side of the core. \mbox{Figure \ref{fig.comparison_saturation_heat_yb_tm}} shows the cuts through the core of the transverse heat profiles.

It is apparent from the heat profiles in Fig.~\ref{fig.comparison_saturation_heat_yb_tm} that the degree of population saturation is significantly higher in the Tm-doped fiber case - note the flatter shapes of the heat profiles throughout the length of the fiber. The total heat deposited in the Tm-doped fiber is about three times larger than deposited in the Yb-doped fiber. The degree of saturation is determined by competition between the rates of pumping into the upper laser level and stimulating emission out of that level. Saturation tends to flatten the heat profile and reduce mode coupling strength\cite{Smith:2013b,Hansen:2014,Tao:2015,Ward:2015}.

\begin{figure}[htb]
\centering
\includegraphics[width=0.9\textwidth]{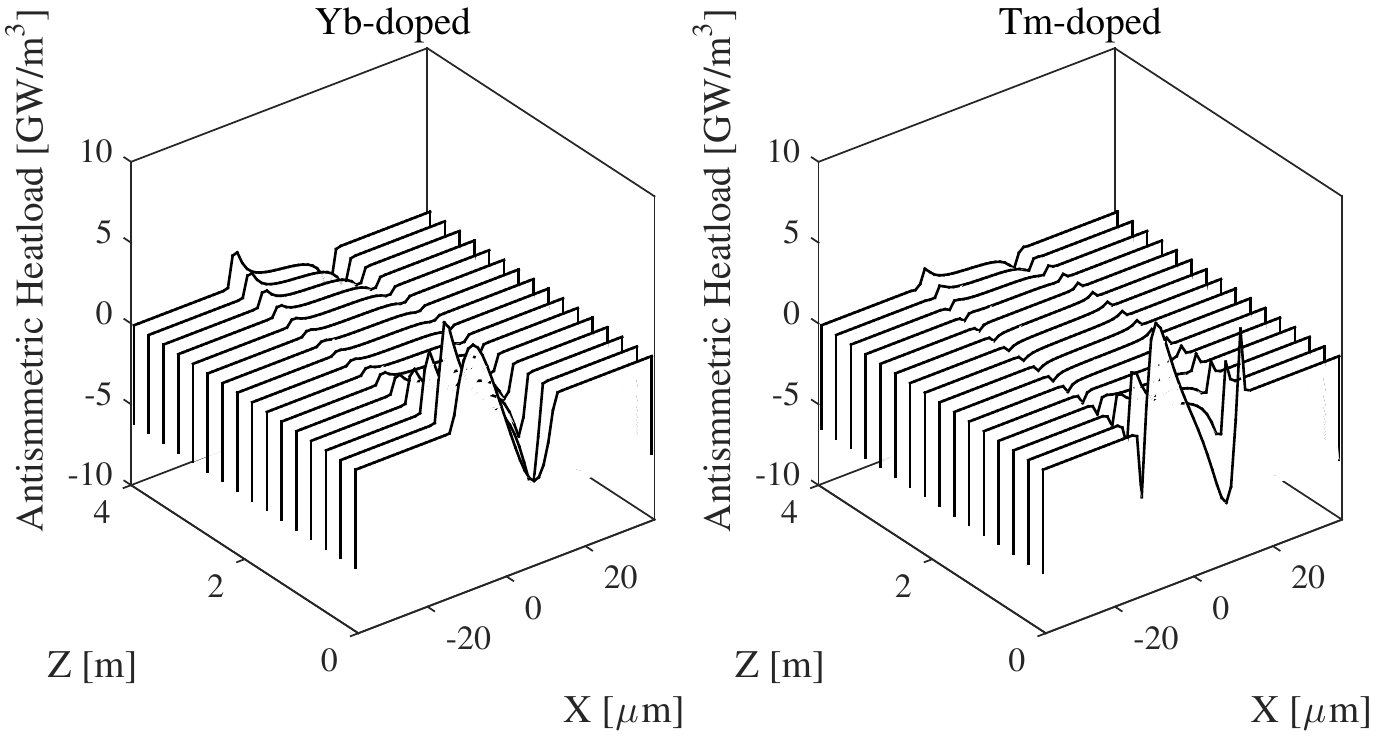}
\caption{Cuts through the fiber center of the total computed antisymmetric parts of the heat profiles for the bidirectionally pumped 25~$\mu$m diameter core Yb- and Tm-doped fibers of Table~\ref{tab.comparison_yb_tm}, with each end of the fiber pumped with 500~W.}
\label{fig.comparison_antisymmetric_saturation_yb_tm}
\end{figure}

Only the antisymmetric part of the heat can create the antisymmetric part of the temperature which is responsible for coupling light between the LP$_{01}$ and LP$_{11}$ modes. The antisymmetric parts of the heat profiles are plotted in Fig.~\ref{fig.comparison_antisymmetric_saturation_yb_tm}. These plots demonstrate that the antisymmetric heating is larger for the Yb-doped fiber than for the Tm-doped fiber despite Yb having one third as much total heat. Although the antisymmetric part of the heat is not a direct measure of the mode coupling gain because the heat must be converted into a temperature profile with an associated phase lag, it still gives a good indication of the relative mode coupling strengths.

\begin{figure}[htb]
\centering
\includegraphics[width=0.9\textwidth]{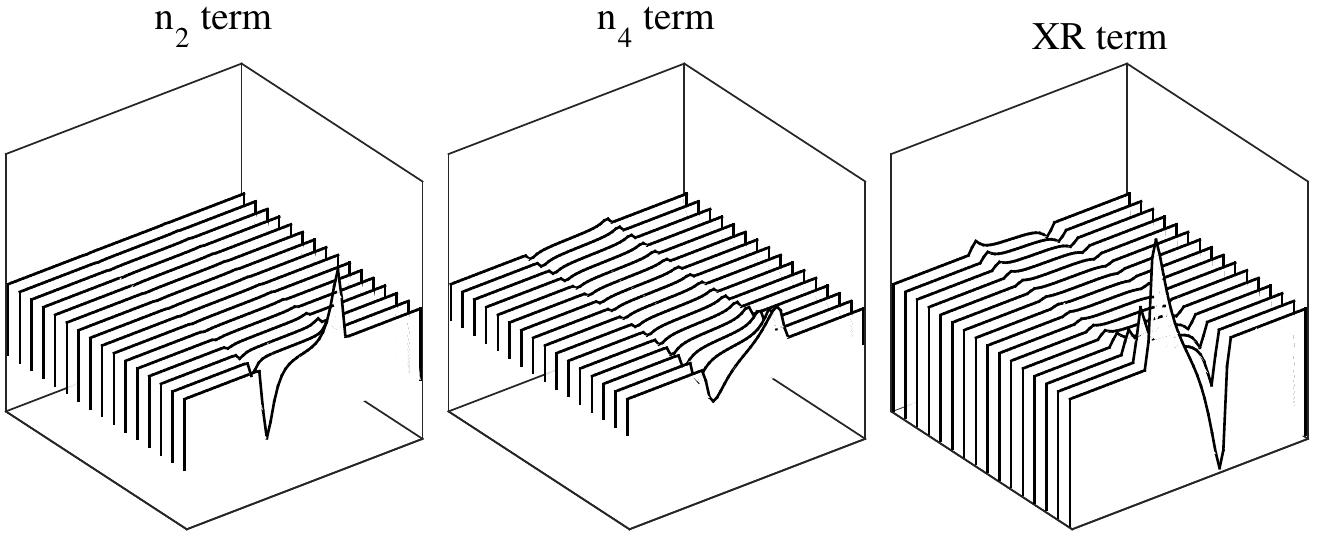}
\caption{Cuts through the fiber center of the computed antisymmetric parts of heating described by the three terms of Eq.~(\ref{eq.heat3}) in the Tm-doped fiber using parameters given in Table~\ref{tab.comparison_yb_tm}. Here, the fiber is bidirectionally pumped with 500 W pump input in each end. The left plot shows heating due to nonradiative decay from level 2; the middle plot shows heating due to nonradiative decay from level 4; the right plot show heating due to the cross relaxation process. The plot axes are the same as in Fig.~\ref{fig.comparison_antisymmetric_saturation_yb_tm}, except the vertical axis ranges from -20 to 20~GW/m$^3$. The sum of these three terms comprises the righthand plot of Fig.~\ref{fig.comparison_antisymmetric_saturation_yb_tm}.}
\label{fig.heat_terms_tm}
\end{figure}

It is interesting to look at the individual contributions to the antisymmetric heat in the Tm-doped fiber. There are three contributions corresponding to the the three terms in Eq.~(\ref{eq.heat3}): one proportional to $n_4$, one proportional to from $n_2$, and one proportional to the cross relaxation rate. In Fig.~\ref{fig.heat_terms_tm} we show these three contributions individually. The cross relaxation contribution has the same sign of asymmetry as the Yb-doped fiber but the other two contributions have the opposite sign. The partial cancellation between the cross relaxation process and nonradiative decay contributions tends to suppress the magnitude of the antisymmetric heating, reducing the STRS gain. 

\section{Stimulated Brillouin suppression}\label{sec:SBS}
Stimulated Brillouin scattering (SBS) is often a problem in high power fiber amplifiers. The SBS threshold is proportional to the effective modal area $A_{\rm eff}$, and decreases with fiber length. This implies that a shorter fiber with larger core should have a higher SBS threshold. However, increasing the core size while keeping the same pump cladding size decreases the ratio of signal to pump irradiance and this tends to decrease the degree of population saturation. This leads to weaker suppression of STRS gain and a reduction in mode instability threshold. This was demonstrated earlier in Yb-doped fibers\cite{Smith:2013b}. To study this effect in Tm-doped fibers, we apply our STRS model to a fiber with doubled core diameter which doubles the V-number for constant NA. This 50~$\mu$m core diameter fiber is 1.2~m in length instead of 4 m, but all other parameters are those of Table~\ref{tab.comparison_yb_tm} (note the pump cladding diameter remains 400~$\mu$m). The frequency of maximum STRS gain changes from 4~kHz to 1.5~kHz due to the increased core diameter, so we use 1.5 kHz as the frequency offset. This short-fat fiber should have an SBS threshold roughly nine times higher than the long-skinny fiber examined above. The question is how large is the STRS threshold reduction?

\begin{table}
\begin{center}
\begin{tabular}{|c|c|c|c|c|c|}
\hline
$L$ & $d_{core}$ & V &    Co-pumped   &    Counter-pumped    &    Bidirectionally-pumped\\
\hline
4~m   & 25 ${\mu}$m & 4.0  	& 2046~W      &     3947~W          &  3210~W  \\
1.2~m & 50 ${\mu}$m & 8.0   & 899~W       &     1244~W          &  1433~W  \\\hline
\multicolumn{3}{|c|}{Ratio} & 2.28 & 3.17 & 2.24 \\
\hline
\end{tabular}
\caption{Signal power at mode instability thresholds for short-fat and long-skinny Tm-doped fiber amplifiers. Ratio is the ratio of long-skinny threshold to short-fat threshold.}
\label{tab.shortfat_longskinny}
\end{center}
\end{table}

Computed mode instability thresholds of the short-fat and long-skinny fiber are compared in Table~\ref{tab.shortfat_longskinny}. The threshold in the short-fat fiber is roughly half that in the long-skinny fiber due to reduced saturation. Thermal lensing is still modest; the maximum reduction of $A_{\rm eff}$ at threshold is about 22\% for copumped case; 32\% for counterpumped; and 24\% for bidirectionally pumped.

\begin{figure}[htb]
\centering
\includegraphics[width=0.9\textwidth]{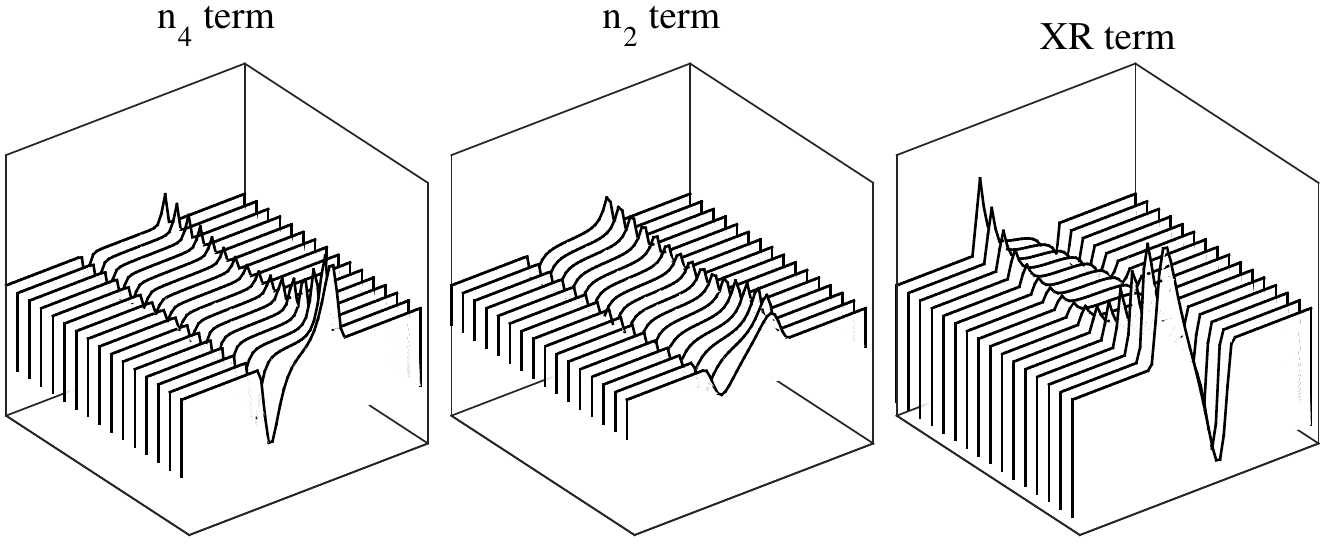}
\caption{Cuts through the fiber center of the computed antisymmetric parts of heating for the 50~$\mu$m core diameter (short-fat) fiber pumped with 500~W from each end. The vertical scale is the same as in Fig.~\ref{fig.heat_terms_tm}. The length scale range is 0 to 1.2~m. The transverse scale range is -75~$\mu$m to 75~$\mu$m. }
\label{fig.heat_terms_tm_shortfat}
\end{figure}

In Fig.~\ref{fig.heat_terms_tm_shortfat} we again look at the contributions from the three heat terms, now for the short-fat fiber pumped with 500~W from each end. The shapes are qualitatively similar to those in Fig.~\ref{fig.heat_terms_tm}, but it's clear the degree of saturation here is much lower. 

\section{Reduced cross relaxation}
As discussed above, our model indicates that a partial cancellation between the cross relaxation and nonradiative decay heat terms contribute to the strong saturation in the Tm-doped fiber. This raises the question whether we can enhance the cancellation between the two terms to raise the threshold still higher. For the short-fat fiber we found that at threshold the antisymmetric part of the heat profile is dominated by the cross-relaxation contribution. For a given population distribution, the strength of the cross-relaxation contribution relative to the nonradiative decay contribution would be decreased by lowering the Tm doping density $N_\circ$. Can this produce stronger cancellation and higher saturation? To answer this, we reduced the doping density and computed the new threshold. We found that contrary to the hope for better cancellation, at threshold the cross-relaxation contribution to the net antisymmetric heat is more dominant at lower doping levels. While the cross-relaxation rate is reduced, the nonradiative decay is reduced more because the population distribution is changed. The result is a reduction in the partial cancellation of the two antisymmetric heating sources and reduction in threshold power.

Our modeled threshold in the bidirectionally-pumped 50 $\mu$m core diameter case (short-fat fiber) described in Section~\ref{sec:SBS} is decreased by 9\% when we decrease Tm doping density by 25\%, and by 22\% when we decrease it by 50\%. Pump absorption is decreased when doping density is reduced, so we modeled fiber lengths of $L=(1.2, 1.6, 2.4)$~m for $N_{\circ} = (5.0, 3.75, 2.5) \times 10^{26}$~m$^{-3}$, which yield power efficiencies at threshold of $(65.7, 63.7, 60.2)\%$. At threshold the total amount of left-behind power is similar for all three cases, at $(744, 739, 762)$~W. We conclude that changing the doping level and with it the rate of cross relaxation does not lead to dramatic changes in the heatload at the instability threshold. The highest threshold occurs at the highest doping density.

\section{Additional features of Tm-doped fiber amplifier}
\subsection{Thermo-optic coefficients}

We have used the thermal properties of silica in our modeling, but the glass used for Tm-doped fibers usually contains a large amount of aluminum. We expect the mode instability gain to scale as approximately $1/({\rm d} n/ {\rm d} T)$. Pure silica has unusually high thermo-optic coefficient among silicate glasses, so aluminum silicate Tm hosts probably have a lower coefficient. This would raise the mode instability threshold above the levels computed here. More deliberate reductions of the thermo-optic coefficient by increasing B or P glass content have also been proposed\cite{Ballato:2014} for the purpose reducing the thermo-optic coefficient and raising the instability threshold.% Tm doped fibers have Al, Ba, Zn, La added in some cases\cite{Lee:2013}.

We found that the predicted mode instability threshold of the copumped short-fat Tm-doped fiber amplifier increased by a factor of 1.8 when we halved the value of ${\rm d}n/{\rm d}T$.

\subsection{Photodarkening}
Photodarkening has been shown to be very important in STRS gain in Yb-doped fiber. A few percent signal power absorption through photodarkening was predicted to reduce the  mode instability threshold by a factor of two \cite{Smith:2013d}, consistent with observed behavior in the laboratory \cite{Jorgensen:2013,Otto:2015}. One particularly interesting difference between Tm-doped fibers and Yb-doped fibers is that photodarkening seems to be absent\cite{Lee:2013,Moulton:2009} in Tm-doped fiber pumped at 795~nm. Photodarkening has been observed in Tm-doped fiber pumped with 1064~nm, ps pulses\cite{Broer:1993}, but it does not seem to occur with 790~nm, CW pumping.

\section{Conclusion}
We have extended our model of STRS in fiber amplifiers to Tm-doped fibers pumped at 790~nm, and found mode instability thresholds to be significantly higher than those of similar Yb-doped fibers despite several times greater heatload. We demonstrated this is due to two effects: a $\lambda$ scaling which accounts for a factor of two due to the approximately doubled signal wavelength in Tm-doped fibers, and an increased degree of population saturation in Tm-doped fibers compared with Yb-doped fibers.  A partial cancellation of the antisymmetric part of the heating between the heat caused by nonradiative decay and caused by the cross-relaxation process in Tm-doped fibers contributes to the gain saturation. %We used the thermo-optic coefficient of silica, but it is likely Tm-doped fibers have reduced thermo-optic coefficient which would further increase the mode instability thresholds.

As was demonstrated earlier for Yb-doped fibers, a balance must be struck between SBS gain and STRS gain in Tm-doped fibers. Raising the SBS threshold tends to reduce saturation and lower STRS thresholds. We also found that the highest STRS thresholds occur at the highest Tm doping levels, where the cross relaxation rate is greatest.

\end{document}